\documentclass{ws-procs_priv}

\usepackage{graphicx}
\newcommand{\gev}{\mbox{\,\rm GeV}}
\newcommand{\tev}{\mbox{\,\rm TeV}}
\newcommand{\met}{$E\!\!\!\!/_{T}$}

\begin{document}
\title{Recent Results from GMSB SUSY Searches at 
  D\O\footnote{\uppercase{T}alk presented by
  \uppercase{T.~C}hristiansen at {\it \uppercase{SUSY} 2003:
  \uppercase{S}upersymmetry in the \uppercase{D}esert}\/, held at the
  \uppercase{U}niversity of \uppercase{A}rizona, \uppercase{T}ucson,
  \uppercase{AZ}, \uppercase{J}une 5-10, 2003. \uppercase{T}o appear
  in the \uppercase{P}roceedings.}\\[-0.5cm]
}

\author{Tim Christiansen \lowercase{for the} D\O~Collaboration}

\address{Ludwig-Maximilians-Universit\"at Munich \\
Am Coulombwall 1\\ 
D-85748 Garching b.~M\"unchen, Germany\\ 
E-mail: Tim.Christiansen@physik.uni-muenchen.de}

%


\maketitle

\abstracts{\vspace*{-0.3cm}
  Using approximately $41\,\rm{pb^{-1}}$ of $p\bar{p}$ collisions,
  recorded with the D\O\ detector during the TeVatron Run II, this
  document describes the search for minimal GMSB SUSY signal (with a
  neutralino as next-to-lightest supersymmetric particle) 
  in $\gamma\gamma+$\met{}\ events. Since no
  deviation from the Standard Model is observed, a $95\,\%$ confidence
  level upper limit on the signal cross-section is calculated.
  This is translated into a lower limit on the neutralino mass,
  yielding  $m(\chi_1^0)>66\gev$ in the context of 
  the {\it Snowmass} GMSB model line\cite{snowmass}.
  }


\vspace*{-0.3cm}
This article describes the search for supersymmetry (SUSY) with
gauge-mediated supersymmetry breaking (GMSB), or --- more generally ---
SUSY with a light gravitino in the framework of the minimal
supersymmetric Standard Model. Supersymmetric models with a
light gravitino ($\tilde{G}$) are characterized by a supersymmetry breaking
scale of the order of $100\,\rm{TeV}$ and a gravitino as the
lightest supersymmetric particle. In these models, the phenomenology
is usually driven by the nature of the next-to-lightest
SUSY particle (NLSP) and by the small coupling to $\tilde{G}$. 
If the NLSP, which is assumed to be the lightest
neutralino ($\chi_1^0$), has a non-vanishing 'Bino' component, it is
unstable and decays into a photon and a gravitino: 
$\chi_1^0\rightarrow\gamma\tilde{G}$. Since the gravitino escapes
detection,
large missing transverse energy (\met) is expected.
%

\vspace*{-0.2cm}
\section*{Search for Supersymmetry in $\gamma\gamma+$\met Events}
Around $41\,\rm{pb}^{-1}$ of $p\bar{p}$ collisions at a center-of-mass
energy of $\sqrt{s}=1.96\,\rm{TeV}$, recorded with the D\O{} detector
between September 2002 and January 2003, are examined for evidence of
GMSB SUSY in $\gamma\gamma+$\met{} events. The data sample was
taken by a combination of triggers  requiring one or more 
highly-energetic electromagnetic clusters in the D\O\ calorimeter.

Photons are reconstructed from energy depositions in the
calorimeters. In order to minimize the background from hadronic 
jets and noise in the calorimeter, the electromagnetic clusters
must have a shower shape consistent with that of a photon and only
little energy is allowed to be detected in the vicinity of the
cluster. The central tracking system must not have a reconstructed
track pointing to the electromagnetic cluster in order to reject
electrons. Only photons with a transverse energy of $E_T>20\gev$ are
considered. The precise calibration of the energy of electromagnetic
showers is achieved using $Z\rightarrow ee$ events.
The missing transverse energy is calculated from the jet and
electromagnetic scale-corrected sum over all calorimeter cells.

All background sources of events with two photons and missing
transverse energy can be divided into two main groups:
The first group, where the missing 
transverse energy is due to mis-measurement, comprises QCD events
with direct photons or hadronic jets, which have been mis-identified as
photons, and Drell-Yan $Z+X\rightarrow ee+X$ processes with both
electrons mis-identified as photons.
The second group are background processes with true missing transverse
energy due to neutrinos which escape detection. This sample is
dominated by $W\gamma\rightarrow e\nu\gamma$ events. 
Additional contributions to this group come from $Wj\rightarrow
e\nu\text{``}\gamma\text{''}$ events (where a hadronic jet is
mis-identified as a photon),
$Z\rightarrow \tau\tau\rightarrow ee+X$ processes, and  $t\bar{t}$
and di-boson production, where the tracks of the final state electrons
are not reconstructed, leading to fake photons.

\begin{figure}[ht]
\vspace{-0.2cm}
\centerline{\epsfxsize=0.70\textwidth 
  \epsfbox{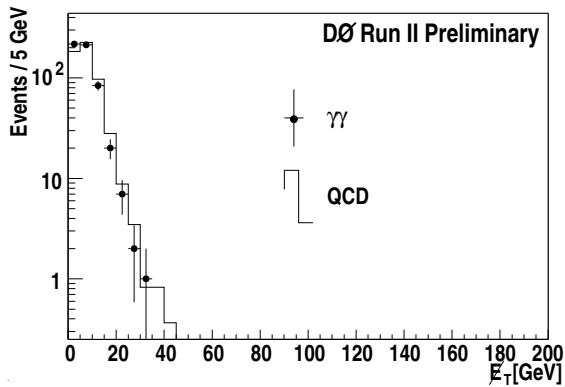}}   
\caption{Missing transverse energy distribution of $\gamma\gamma$ events and
the QCD background. The QCD background is normalized to the
$\gamma\gamma$ sample at \met{}$<20\gev$.\label{fig:met}}
\end{figure}

A QCD-enriched event sample is obtained by inverting the
shower-shape criterion for at least one of the two energy
clusters. For these events, the \met{}-resolution is expected to be
very similar for events with real photons and with jets faking photons,
since a jet, that mimics a photon, fragments into one or more neutral
pions. 
The QCD sample is normalized to the
di-photon sample at small transverse energies, 
\met{}$< 20\gev$. Figure \ref{fig:met} shows the \met{}-distribution for the
$\gamma\gamma$ sample and the QCD background.

All Standard-Model backgrounds with true \met{} involve electrons and
not photons. This contribution is measured with an electron-photon
sample for which the QCD portion is subtracted in the same way as
described above. The normalization of the electron-photon sample is
determined from the measured probabilities of identifying an electron as
an electron (photon). Table \ref{tab:nevts} shows the numbers of
remaining events for various cuts on the missing transverse energy.

\begin{table}[ht]
\tbl{Number of $\gamma\gamma$ and background events for various cuts
  on \met.}
{\footnotesize
\begin{tabular}{@{}lrrr@{}}
\hline
{} &{} &{} &{} \\[-1.5ex]
{} & \met{}$>25\gev$ & \met{}$>30\gev$ & \met{}$>35\gev$ \\[1ex]
\hline
{} &{} &{} &{} \\[-1.5ex]
$\gamma\gamma$ events & $3$ & $1$ & $0$ \\[1ex]
\hline
{} &{} &{} &{} \\[-1.5ex]
QCD (with wrong \met)   & $6.0\pm 0.8$ & $2.5\pm 0.5$ & $1.6\pm 0.4$ \\[1ex]
$e+\nu+\gamma/j$ events & $0.6\pm 0.4$ & $0.2\pm 0.2$ & $0.0\pm 0.2$ \\[1ex]
\hline
{} &{} &{} &{} \\
\end{tabular}\label{tab:nevts} }
\vspace*{-13pt}
\end{table}

The Monte-Carlo events for the SUSY signal are processed through a
full detector simulation and reconstructed the same way as the data. 
The SUSY signal has been generated for various points on the minimal-GMSB
{\it Snowmass} slope\cite{snowmass} with a neutralino NLSP.
This model line has one free parameter, $\Lambda$, which
determines the scale of SUSY breaking. The minimal GMSB parameters
of this model line are the messenger mass scale $M=2\Lambda$, the number of
messenger fields $N_5=1$, the ration of Higgs vacuum expectation values
$\tan{\beta}=15$, and the sign of the Higgsino mass term $\mu>0$.

While each signal point has its own optimum cut on \met, a common
cut at \met{} $> 30\gev$ is chosen. The signal acceptance varies
between $3\,\%$ and $17\,\%$ for the mass points between $\Lambda=35\tev$
and $\Lambda=60\tev$, respectively. 

\section*{Conclusion}
\begin{figure}[ht]
\centerline{\epsfxsize=0.75\textwidth 
  \epsfbox{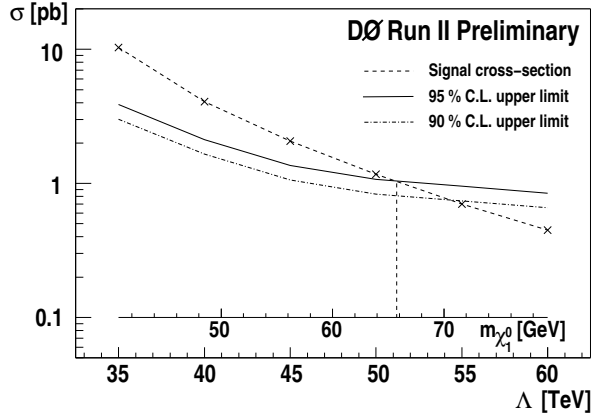}}
\vspace{-0.4cm}   
\caption{\label{fig:limit}
$95\,\%$ confidence level upper cross-section limits for the Snowmass
  model line for minimal GMSB SUSY with a neutralino as NLSP as
  compared to theoretical predictions.}
\end{figure}

Since no excess of $\gamma\gamma$ events is observed, $95\,\%$
confidence level (C.L.) upper cross-section limits are calculated and
compared to the theoretical predictions (see Figure
\ref{fig:limit}). The $95\,\%$ C.L.\ lower limit on the parameter
$\Lambda$ is $51\tev$, corresponding to lightest neutralino and
chargino masses of $66\gev$ and $116\gev$, respectively. The
Run I limits in this channel using {\it similar} models were 
$m(\chi_1^0)>75\gev$ in D\O\cite{D0RunI} 
and $m(\chi_1^0)>65\gev$ in CDF\cite{CDFRunI}.

The analysis described in this document has been updated by the 
D\O\ collaboration for the summer conferences in 2003, using data
recorded between August 2002 and June 2003:
This latest D\O~result in the $\gamma\gamma+$\met{} channel,
which is based on an integrated luminosity of $(128\pm 13)\,\rm{pb^{-1}}$, 
yields a $95\,\%$ C.L.\ lower limit of $M(\chi_1^0)>80\gev$ on the 
lightest-neutralino mass. 
%
%
%
With more and more data being collected at D\O~Run II and with
additional complementary analyses (see for instance reference
\cite{Culbertson:2000am}), the D\O~collaboration will be able to probe
for SUSY signal in the light of GMSB from various angles and it can be
expected that the sensitivity reaches far into regions of the SUSY
parameter space which have not been excluded yet.


%



\end{document}